# Effect of stress induced anisotropy and applied magnetic field on second order perturbed energy of thin ferromagnetic films

P. Samarasekara

Department of Physics, University of Peradeniya, Peradeniya, Sri Lanka

## Abstract

All the simulations were carried out for sc(001) film with three (N=3) layers. The effect of applied magnetic field and stress induced anisotropy on energy of ultra-thin ferromagnetic films has been investigated using Heisenberg Hamiltonian with second order perturbation. The energy becomes minimums at certain values of angles, applied magnetic field and stress, by indicating that film can be easily oriented in theses certain directions by applying particular applied magnetic field or stress. For example, when applied magnetic field ($\frac{H_{out}}{\omega}$) perpendicular to film plane is 3.6, the film can be easily oriented along θ=0.97 and θ=2.65 directions. When in plane applied magnetic field is $\frac{H_{in}}{\omega}$ =3, energy minimums can be detected at θ=1 and θ=2.6 radians. When applied stress ($\frac{K_s}{\omega}$) is 2.7, film can be easily oriented in θ=2.8 direction.

## 1. Introduction:

The theories of exchange anisotropy have been extensively investigated, because of the difficulties of understanding the behavior of exchange anisotropy and its applications in magnetic sensors and media technology [2]. Due to their potential applications of magnetic memory devices and microwave devices, ferromagnetic films are thoroughly studied nowadays. Bloch spin wave theory has been applied to study magnetic properties of ferromagnetic thin films earlier [3]. The magnetization of some thin films is oriented in plane due to dipole interaction. But the perpendicular orientation is preferred at the surface due to the broken symmetry of uniaxial anisotropy energy. Two dimensional Heisenberg model has been earlier used to explain the magnetic anisotropy in the presence of dipole interaction [4]. Ising model has been used to study magnetic properties of ferromagnetic thin films with alternating super layers [5].

When the film is non-oriented the effect of second order perturbations has to be taken into account [9, 11]. The stress induced anisotropy plays an important role in the explanation of Magnetic properties of materials with low crystal anisotropy such as ferrites [6, 7, 8, 12]. Internally induced stress of a thin film mainly arises in the cooling or heating process of the thin film, due to the difference between the thermal expansion coefficients of substrate and the film. The demagnetization factor ($N_d$) described here is 1 and 0 in perpendicular direction and within the film plane in SI units, respectively. Previously the variation of energy in ultra-thin ferromagnetic films with two and three layers has been studied using second order perturbation of Heisenberg Hamiltonian with magnetic exchange interaction, second order anisotropy and stress induced terms only [9]. Easy and hard directions of sc(001), fcc(001) and bcc(001) lattices with the effect of second order anisotropy only are $34.4^0$ and $124.4^0$, respectively. The angle between easy and hard directions was exactly $90^0$ as expected [9]. Second and fourth order anisotropy constants were assumed to be unchanged for all layers of the film. But all energy terms for a film with three layers will be taken into account for the energy calculations given in this report. The energy of ferromagnetic and ferrite has been earlier investigated in detail [10, 13, 14, 15, 16, 17].

## 2. Model:

According to the Heisenberg Hamiltonian given in previous paper [9], the total energy can be given as following.

$$E(\theta) = E_0 + \vec{\alpha}.\vec{\varepsilon} + \frac{1}{2}\vec{\varepsilon}.C.\vec{\varepsilon} = E_0 - \frac{1}{2}\vec{\alpha}.C^+.\vec{\alpha}$$

Here matrix elements of matrix C are given by

$$C_{mn} = -(JZ_{|m-n|} - \frac{\omega}{4}\Phi_{|m-n|}) - \frac{3\omega}{4}\cos 2\theta \Phi_{|m-n|} + \frac{2N_d}{\mu_0}$$

$$+ \delta_{mn}\{\sum_{\lambda=1}^{N}[JZ_{|m-\lambda|} - \Phi_{|m-\lambda|}(\frac{\omega}{4} + \frac{3\omega}{4}\cos 2\theta)] - 2(\sin^2\theta - \cos^2\theta)D_m^{(2)}$$

$$+ 4\cos^2\theta(\cos^2\theta - 3\sin^2\theta)D_m^{(4)} + H_{in}\sin\theta + H_{out}\cos\theta - \frac{4N_d}{\mu_0} + 4K_s \sin 2\theta\}$$

and $\vec{\alpha}(\varepsilon) = \vec{B}(\theta)\sin 2\theta$ are the terms of matrices with

$$B_\lambda(\theta) = -\frac{3\omega}{4}\sum_{m=1}^{N}\Phi_{|\lambda-m|} + D_\lambda^{(2)} + 2D_\lambda^{(4)}\cos^2\theta \qquad (1)$$

Here J, $Z_{|m-n|}$, $\omega$, $\Phi_{|m-n|}$, $\theta$, $D_m^{(2)}, D_m^{(4)}, H_{in}, H_{out}, N_d, K_s$, m, n and N are spin exchange interaction, number of nearest spin neighbors, strength of long range dipole interaction, constants for partial summation of dipole interaction, azimuthal angle of spin, second and fourth order anisotropy constants, in plane and out of plane applied magnetic fields, demagnetization factor, stress induced anisotropy constant, spin plane indices and total number of layers in film, respectively. When the stress applies normal to the film plane, the angle between $m^{th}$ spin and the stress is $\theta_m$. $E_0$ is the energy of the oriented thin ferromagnetic film.

The matrix elements for a film with three layers can be given as following [9],

$$C_{12} = C_{21} = C_{23} = C_{32} = -JZ_1 + \frac{\omega}{4}\Phi_1(1 - 3\cos 2\theta) + \frac{2N_d}{\mu_0}$$

$$C_{13} = C_{31} = -JZ_2 + \frac{\omega}{4}\Phi_2(1 - 3\cos 2\theta) + \frac{2N_d}{\mu_0}$$

$$C_{11} = C_{33} = J(Z_1 + Z_2) - \frac{\omega}{4}(\Phi_1 + \Phi_2)(1 + 3\cos 2\theta) - \frac{2N_d}{\mu_0} + (2\cos 2\theta)D_m^{(2)}$$

$$+ 4\cos^2\theta(\cos^2\theta - 3\sin^2\theta)D_m^{(4)} + H_{in}\sin\theta + H_{out}\cos\theta + 4K_s \sin 2\theta$$

$$C_{22} = 2JZ_1 - \frac{\omega}{2}\Phi_1(1 + 3\cos 2\theta) - \frac{2N_d}{\mu_0} + (2\cos 2\theta)D_m^{(2)}$$

$$+ 4\cos^2\theta(\cos^2\theta - 3\sin^2\theta)D_m^{(4)} + H_{in}\sin\theta + H_{out}\cos\theta + 4K_s \sin 2\theta$$

When the second or fourth order anisotropy constants do not change inside an ultra thin film, $D_1^{(2)}=D_2^{(2)}=D_3^{(2)}$ and $D_1^{(4)}=D_2^{(4)}=D_3^{(4)}$. Under some special conditions [9], $C^+$ is the standard inverse of a matrix, given by matrix element $C^+_{mn} = \dfrac{cofactor C_{nm}}{\det C}$. Therefore for the convenience the matrix elements $C^+_{mn}$ will be given in terms of $C_{11}$, $C_{22}$, $C_{32}$, and $C_{31}$ only.

$$C^+_{11} = \frac{C_{11}C_{22} - C_{32}^2}{C_{11}(C_{11}C_{22} - C_{31}^2) + 2C_{32}^2(C_{31} - C_{11})} = C^+_{33}$$

$$C^+_{12} = \frac{C_{32}C_{31} - C_{32}C_{11}}{C_{11}(C_{11}C_{22} - C_{31}^2) + 2C_{32}^2(C_{31} - C_{11})} = C^+_{21} = C^+_{23} = C^+_{32}$$

$$C^+_{13} = \frac{C_{32}^2 - C_{22}C_{31}}{C_{11}(C_{11}C_{22} - C_{31}^2) + 2C_{32}^2(C_{31} - C_{11})} = C^+_{31}$$

$$C^+_{22} = \frac{C_{11}^2 - C_{31}^2}{C_{11}(C_{11}C_{22} - C_{31}^2) + 2C_{32}^2(C_{31} - C_{11})} \quad (2)$$

Both matrices C and $C^+$ are highly symmetric, and total energy can be given as [9],

$E(\theta)=E_0-0.5[C^+_{11}(\alpha_1^2+\alpha_3^2)+C^+_{32}(2\alpha_1\alpha_2+2\alpha_2\alpha_3)+C^+_{31}(2\alpha_1\alpha_3)+\alpha_2^2 C^+_{22}]$

From equation 1,

$$B_1(\theta) = B_3(\theta) = -\frac{3\omega}{4}(\Phi_0 + \Phi_1 + \Phi_2) + D_\lambda^{(2)} + 2D_\lambda^{(4)} \cos^2\theta$$

$$B_2(\theta) = -\frac{3\omega}{4}(\Phi_0 + 2\Phi_1) + D_\lambda^{(2)} + 2D_\lambda^{(4)} \cos^2\theta$$

Because in this case, $\alpha_1=\alpha_3$

$E(\theta)=E_0-0.5[2C^+_{11}\alpha_1^2+4C^+_{32}\alpha_1\alpha_2+2C^+_{31}\alpha_1^2+\alpha_2^2 C^+_{22}]$  (3)

Here [9]

$$E_0 = -\frac{J}{2}[NZ_0 + 2(N-1)Z_1] + \{N\Phi_0 + 2(N-1)\Phi_1\}(\frac{\omega}{8} + \frac{3\omega}{8}\cos 2\theta)$$

$$- N(\cos^2\theta D_m^{(2)} + \cos^4\theta D_m^{(4)} + H_{in}\sin\theta + H_{out}\cos\theta - \frac{N_d}{\mu_0} + K_s \sin 2\theta)$$

This simulation will be carried out for

$$\frac{J}{\omega} = \frac{D_m^{(2)}}{\omega} = \frac{N_d}{\mu_0 \omega} = \frac{K_s}{\omega} = 10, \frac{D_m^{(4)}}{\omega} = 5 \text{ and } H_{in}=0$$

For sc(001) lattice, $Z_0=4$, $Z_1=1$, $Z_2=0$, $\Phi_0=9.0336$, $\Phi_1= -0.3275$ and $\Phi_2=0$ [1],

$$\frac{C_{12}}{\omega} = \frac{C_{21}}{\omega} = \frac{C_{23}}{\omega} = \frac{C_{32}}{\omega} = 9.92 + 0.2456\cos 2\theta$$

$$\frac{C_{13}}{\omega} = \frac{C_{31}}{\omega} = 20$$

$$\frac{C_{11}}{\omega} = \frac{C_{33}}{\omega} = -9.92 + 20.2456\cos 2\theta + 20\cos^2\theta(\cos^2\theta - 3\sin^2\theta) + \frac{H_{out}}{\omega}\cos\theta + 40\sin 2\theta$$

$$\frac{C_{22}}{\omega} = 0.164 + 20.49\cos 2\theta + 20\cos^2\theta(\cos^2\theta - 3\sin^2\theta) + \frac{H_{out}}{\omega}\cos\theta + 40\sin 2\theta$$

$$\frac{\alpha_1}{\omega} = \frac{\alpha_3}{\omega} = (3.47 + 10\cos^2\theta)\sin 2\theta$$

$$\frac{\alpha_2}{\omega} = (3.716 + 10\cos^2\theta)\sin 2\theta$$

$$\frac{E_0}{\omega} = 25.22 + 9.67\cos 2\theta - 3(10\cos^2\theta + 5\cos^4\theta + \frac{H_{out}}{\omega}\cos\theta + 10\sin 2\theta)$$

### 3. Results and discussion:

Matrix elements $C^+_{mn}$ can be found from equation 2, and hence final energy can be found from equation 3. The 3-D plot of energy versus angle (in radians) and $\frac{H_{out}}{\omega}$ is given in figure 1. Several energy minimums can be observed in this graph indicating that the film can be easily oriented in some directions by applying some external magnetic field perpendicular to film plane. For example, at $\frac{H_{out}}{\omega}=3.6$ the film can be easily oriented in one particular direction. The other easy directions corresponding to this magnetic field $\frac{H_{out}}{\omega}=3.6$ can be found by plotting the 2-D graph between angle and energy. This graph is given in figure 2. The film can be easily oriented along $\theta=0.97$ and $\theta=2.65$ directions under the influence of this magnetic field.

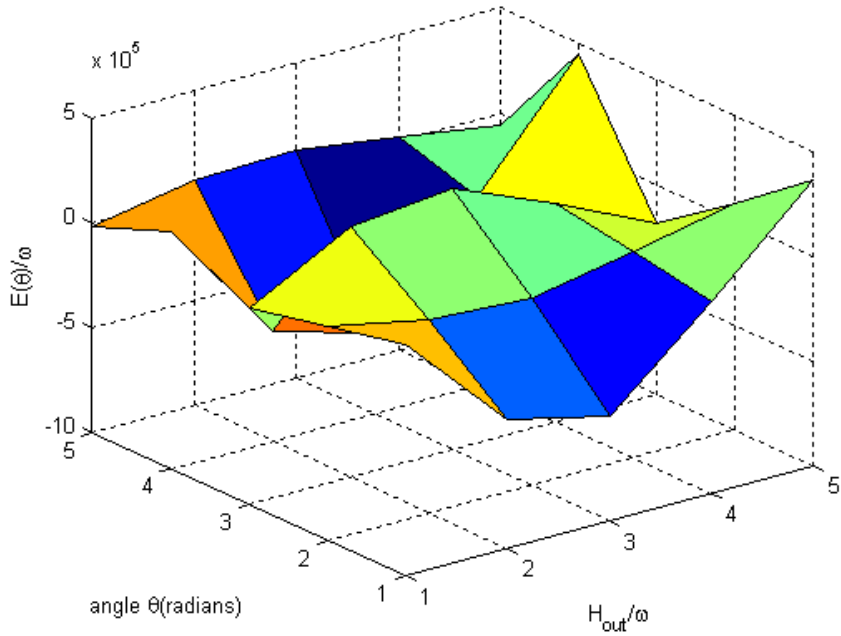

Figure 1: 3-D plot of energy versus angle and applied magnetic field ($\frac{H_{out}}{\omega}$)

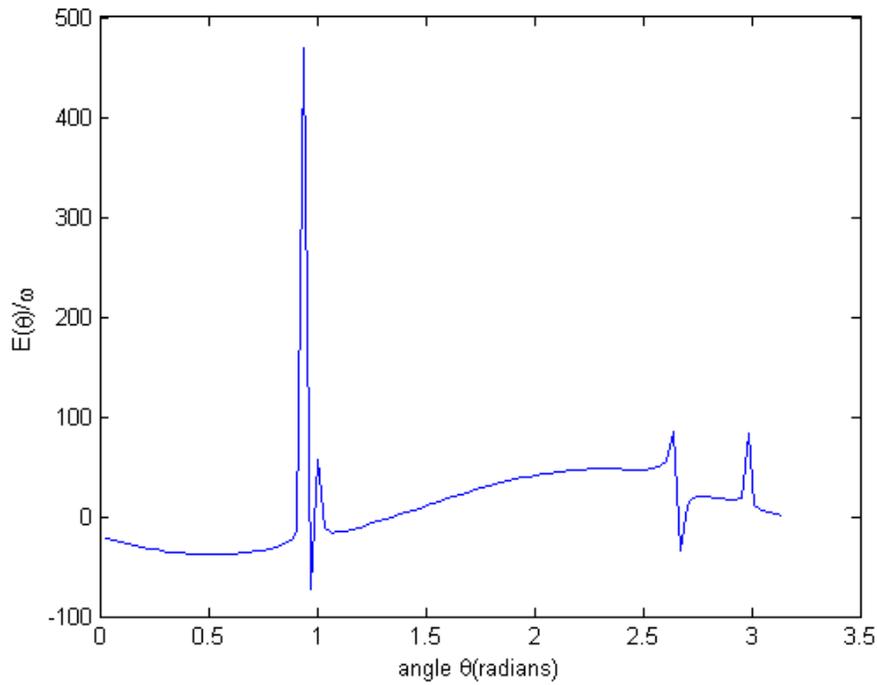

Figure 2: Graph between energy and angle at $\frac{H_{out}}{\omega}=3.6$

When $\dfrac{J}{\omega} = \dfrac{D_m^{(2)}}{\omega} = \dfrac{N_d}{\mu_0 \omega} = \dfrac{K_s}{\omega} = 10, \dfrac{D_m^{(4)}}{\omega} = 5$ and $H_{out}=0$

$$\dfrac{C_{11}}{\omega} = \dfrac{C_{33}}{\omega} = -9.92 + 20.2456\cos 2\theta + 20\cos^2\theta(\cos^2\theta - 3\sin^2\theta) + \dfrac{H_{in}}{\omega}\sin\theta + 40\sin 2\theta$$

$$\dfrac{C_{22}}{\omega} = 0.164 + 20.49\cos 2\theta + 20\cos^2\theta(\cos^2\theta - 3\sin^2\theta) + \dfrac{H_{in}}{\omega}\sin\theta + 40\sin 2\theta$$

$$\dfrac{E_0}{\omega} = 25.22 + 9.67\cos 2\theta - 3(10\cos^2\theta + 5\cos^4\theta + \dfrac{H_{in}}{\omega}\sin\theta + 10\sin 2\theta)$$

All other equations in this case are exactly same as the equation given in previous case. 3-D graph of energy versus angle and $\dfrac{H_{in}}{\omega}$ is given in figure 3. Similar to previous graph, several energy minimums can be observed in this 3-D plot indicating that film can be easily oriented along these directions by applying an external in plane magnetic field. Lowest energy minimum is at $\dfrac{H_{in}}{\omega}$ =3. The graph between energy and angle was plotted to find the easy direction at this applied field as shown in figure 4. Energy minimums can be seen at θ=1 and θ=2.6. But the energy corresponding to minimum in this case is lower than that of previous case.

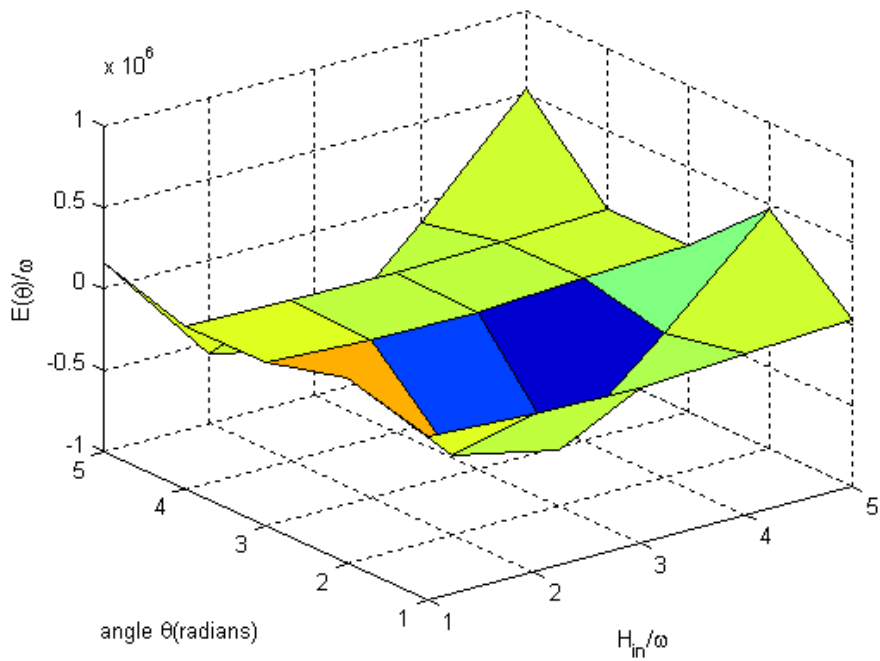

Figure 3: 3-D graph of energy versus angle and applied magnetic field ($\frac{H_{in}}{\omega}$)

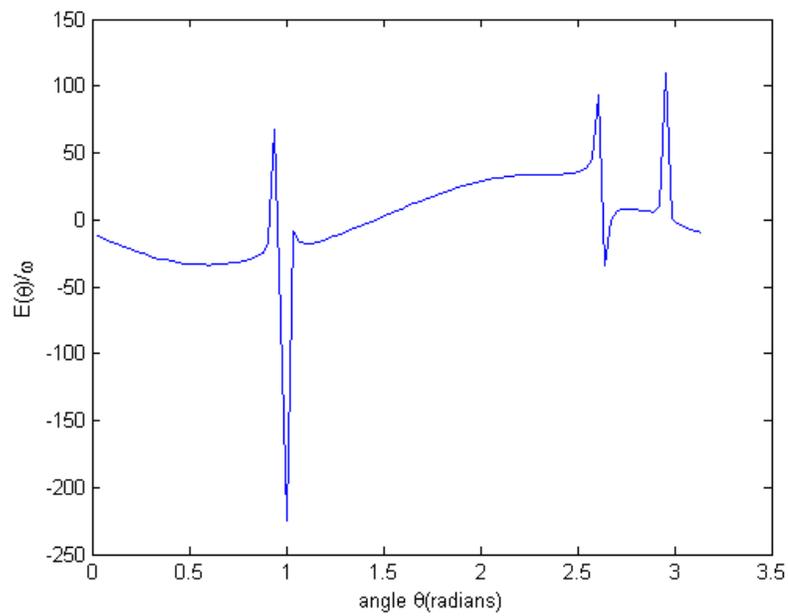

Figure 4: Graph between energy and angle at $\frac{H_{in}}{\omega} = 3$

When $\dfrac{J}{\omega} = \dfrac{D_m^{(2)}}{\omega} = \dfrac{H_{in}}{\omega} = \dfrac{H_{out}}{\omega} = \dfrac{N_d}{\mu_0 \omega} = 10, and\ \dfrac{D_m^{(4)}}{\omega} = 5$

3-D plot of energy versus stress ($\dfrac{K_s}{\omega}$) and angle is given in figure 5. According to this graph, film can be easily oriented on some directions by applying some certain stress perpendicular to the film plane. When applied stress is $\dfrac{K_s}{\omega}$=2.7, film can be easily oriented in θ=2.8 direction.

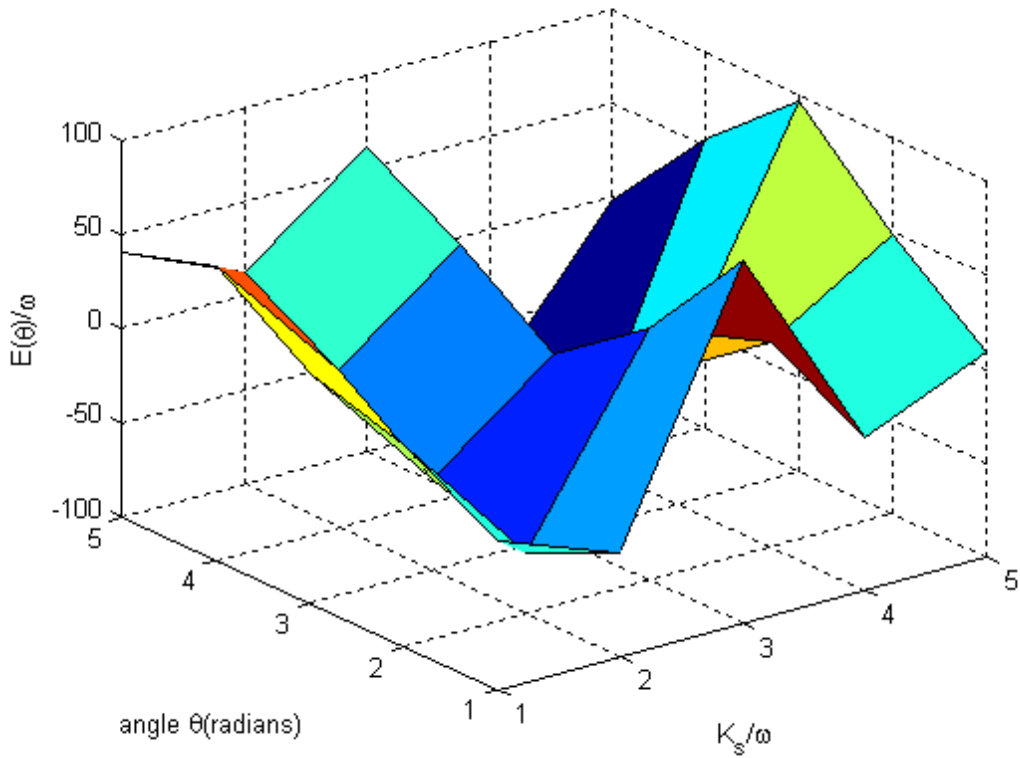

Figure 5: 3-D plot of energy versus stress ($\dfrac{K_s}{\omega}$) and angle

## 4. Conclusion:

The energy becomes minimums at certain values of angles, applied magnetic field and stress, by indicating that film can be easily oriented in theses certain directions by applying particular applied magnetic field or stress. If magnetic field $\frac{H_{out}}{\omega}$ = 3.6 is applied on sc(001) film with three layers, the easy directions of film can be θ=0.97 or θ=2.65 radians. When in plane applied magnetic field is $\frac{H_{in}}{\omega}$ =3, energy minimums can be observed at θ=1 and θ=2.6 radians for sc(001) film with three layers. Under the influence of applied stress $\frac{K_s}{\omega}$ = 2.7, film can be easily oriented in direction of θ=2.8 radians. Although this simulation has been performed for certain values of energy terms, this simulation can be carried out for any other values of energy terms as well.